\documentclass[aps,prd,showkeys,nofootinbib,superscriptaddress,11pt]{revtex4-2}

\usepackage{color}
\definecolor{darkgreen}{rgb}{0,0.35,0}

\usepackage{hyperref}
\usepackage{amsfonts}
\usepackage{amsmath}
\usepackage{bbold}
\usepackage[noabbrev]{cleveref}
\usepackage{subfig}
\usepackage{mathtools,slashed,tensor}
\usepackage[font=scriptsize,labelfont=bf]{caption}
\usepackage{float}
\usepackage[top=2cm, bottom=2cm, left=3cm, right=3cm]{geometry}
\usepackage{tikz-cd}
\usepackage{siunitx}

\newcommand{\be}{\begin{equation}}

\newcommand{\ee}{\end{equation}}
\newcommand{\bea}{\begin{eqnarray}}
\newcommand{\eea}{\end{eqnarray}}
\newcommand{\bef}{\begin{figure}}
\newcommand{\eef}{\end{figure}}
\newcommand{\bce}{\begin{center}}
\newcommand{\ece}{\end{center}}
\newcommand{\ii}{\ensuremath{\mathrm{i}}}
\def\lsim{\mathrel{\rlap{\lower4pt\hbox{\hskip1pt$\sim$}}
    \raise1pt\hbox{$<$}}}         
\def\gsim{\mathrel{\rlap{\lower4pt\hbox{\hskip1pt$\sim$}}
    \raise1pt\hbox{$>$}}}         

\newcommand{\ucharles}{Faculty of Mathematics and Physics, Charles University, V Hole\v{s}ovi\v{c}k\'ach 2, 18000 Prague 8, Czech Republic}
\newcommand{\uach}{Instituto de Ciencias F\'isicas y Matem\'aticas, Universidad Austral de Chile, Casilla 567, 5090000 Valdivia, Chile}
\newcommand{\infn}{Dipartimento di Matematica e Informatica, via Ospedale 72, 09123, Cagliari, Italy and INFN Sezione di Cagliari, Italy}
\newcommand{\borisaff}{Wohlergasse 6, 1100 Vienna, Austria}

\begin{document}
\title{Shadows of new physics on Dirac materials, \\ 
analog GUPs and other amusements}

\author{Alfredo Iorio}
\email{alfredo.iorio@mff.cuni.cz}
\affiliation{\ucharles}
\author{Boris Iveti\'{c}}
\email{bivetic@yahoo.com}
\affiliation{\borisaff}
\author{Salvatore Mignemi}
\email{smignemi@unica.it}
\affiliation{\infn}
\author{Pablo Pais}
\email{pais@ipnp.troja.mff.cuni.cz}
\affiliation{\uach}
\affiliation{\ucharles}

\begin{abstract}
We discuss here how, when higher-order effects in the parameter $\frac{\ell}{\hbar} |\vec{p}|$, related to the lattice spacing $\ell$, are considered, pristine graphene, and other Dirac materials, can be used as tabletop systems where generalized commutation relations are naturally realized. Such generalized algebras of quantization, which lead to generalized versions of the Heisenberg uncertainty principle, are under intense scrutiny these days, as they could manifest a fundamental length scale of spacetime. Despite the efforts and the many intriguing results, there are no experimental signatures of any generalized uncertainty principle (GUP). Therefore, our results here, which tell how to use tabletop physical systems to test certain GUPs in analog experiments, should be of interest to practitioners of quantum gravity. We identify three different energy regimes that we call ``layers'', where the physics is still of a Dirac type but within precisely described limits. The higher the energy, the more sensitive the Dirac system becomes to the effects of the lattice. Here such lattice plays the role of a discrete space where the Dirac quasi-particles live. With the goals just illustrated, we had to identify the mapping between the high-energy coordinates, $X^i$, and the low-energy ones, $x^i$, i.e., those measured in the lab. We then obtained three generalized Heisenberg algebras. For two of them we have the noticeable result that $X^i = x^i$, and for the third one we obtained an improvement with respect to an earlier work: the generalized coordinates expressed in terms of the standard phase space variables, $X^i(x,p)$, and higher order terms.
\end{abstract}

\maketitle

\section{Introduction}
\label{intro}

In the search for new physics, it is an established tradition to look for effects of high energy phenomena at our low energy scales. One way to do it is to consider the action for the ``high-energy'' fields, $\hat{\Phi}_a$, as an expansion in terms of ``low-energy'' fields, $\Phi_a$ and a suitable parameter $\epsilon$, so that
\begin{eqnarray} \label{hepSlowS}
{\cal S} (\hat{\Phi}_a) = {\cal S} (\Phi_a) + S_1 (\epsilon, \Phi_a, \partial) +  O(\epsilon^2,\partial^2) \,.
\end{eqnarray}
Examples from earlier research are many. Two that come to mind are the noncommutative field theories faced with the help of the Seiberg-Witten map \cite{Nathan_Seiberg_1999}
\begin{eqnarray}\label{SWmap}
  \hat{A}_\mu & = & A_\mu - \frac{1}{2} \theta^{\alpha \beta}\,A_\alpha\,(\partial_\beta A_\mu + F_{\beta \mu}) + O(\theta^2) \;, \nonumber \\
  \hat{F}_{\mu\nu} & = & F_{\mu\nu} + \theta^{\alpha \beta}\,(F_{\mu \alpha}\,F_{\nu \beta} - A_{\alpha}\,\partial_{\beta}F_{\mu\nu}) +  O(\theta^2) \;,
\end{eqnarray}
that for (\ref{hepSlowS}) gives
\begin{equation}\label{NCMaxwell}
{\cal S} (\hat{A}_\mu) \ = \ - \frac{1}{4} \hat{F} \cdot \hat{F} \ = \ - \frac{1}{4} \left( F \cdot F - \frac{1}{2} ( \theta \cdot F) \, (F \cdot F)
  +  2 (F\,\theta\,F) \cdot F  \right) + O(\theta^2) \,,
\end{equation}
where $\theta \sim {\ell}_P^2$ and $\ell_P$ is a reference length, usually taken to be the Planck length. For more on this see \cite{Nathan_Seiberg_1999,Madore2000-bj,iorio2002space,iorio2008comment} for the theory and, e.g., \cite{Guralnik2001450,JACKIW200230,castorina2004noncommutative} for some phenomenology.

Another, somehow more general, example is Colladay-Kostelecky's Standard Model Extension \cite{Colladay_Kostelecky1998}, that can be described by
\begin{eqnarray}
{\cal S} (\hat{\Phi}_a) = {\cal S}_{SM} (\Phi_a) + \sum_{k=1}^{\infty} C^{(k)}_{\mu ... \nu} (\Phi_a, \partial)^{\mu ... \nu} \,,
\end{eqnarray}
with $C^{(k)} \sim {\ell}_P^k$. For a review of both approaches, and a comparison, see, e.g., \cite{dice2006}.

These are the days of the generalized uncertainty principles (GUPs) \cite{Maggiore:1993rv,Kempf:1993bq,Scardigli:1999jh}; see also the recent review \cite{FabioEternity}. Many approaches exist that consider high-energy phase space variables, $(x,p)$, vs low-energy variables, $(x_0,p_0)$. Here we focus on a specific one that has been introduced by Ali, Das and Vagenas (ADV) in \cite{AliDasVagenas1,AliDasVagenas2,AliDasVagenas3,AliDasVagenas4}
\begin{eqnarray}
p_i = p_{0 i} \left( 1 - A |\vec{p_0}| + 2 A^2 |\vec{p_0}|^2 \right) \,,
\end{eqnarray}
where $A = \tilde{A} \, \ell_P / \hbar$, $\tilde{A}$ is a numerical factor and
\begin{eqnarray}
x_i = x_{0 i} \,,
\end{eqnarray}
with $[x_{0 i}, p_{0 j}] = \imath \hbar \delta_{i j}$, but where
\begin{eqnarray}\label{ADVGUPtrue}
[x_i,p_j] = \imath \hbar \left[ \delta_{i j} - A |\vec{p}| \left( \delta_{i j} + \frac{p_i p_j}{|\vec{p}|^2} \right) + A^2 |\vec{p}|^2 \left(\delta_{i j} + 3 \, \frac{p_i p_j}{|\vec{p}|^2}  \right) \right] \,.
\end{eqnarray}

There is an intense activity to find GUP-corrected physics. Besides the already cited work of \cite{AliDasVagenas1,AliDasVagenas2,AliDasVagenas3,AliDasVagenas4}, other important results that use different approaches are in \cite{Buoninfante:2019fwr,Petruzziello:2020wkd,Bosso:2021koi}. A compelling example for this paper, for instance, is in \cite{AliDasVagenas4} where it is found that the GUP-corrected Dirac equation, in any dimension, is
\begin{eqnarray}
H \, \psi & = & \left( c \vec{\alpha} \cdot \vec{p} + \beta m c^2 \right) \, \psi \nonumber \\
& = & \left( c \vec{\alpha} \cdot \vec{p}_0 - c A (\vec{\alpha} \cdot \vec{p}_0)^2 + \beta m c^2 \right) \, \psi  \\
& = & E \, \psi \,,\nonumber
\end{eqnarray}
that in two dimensions gives
\begin{eqnarray}
\left(  - \imath \hbar c \sigma_1 \frac{d}{d x_0} + A \, \hbar^2 c \, \frac{d^2}{d x_0^2} + \beta m c^2 \right) \, \psi &=& E \, \psi \,.
\end{eqnarray}

This search for what we might call the ``shadows of new physics'' at our energy scales is fascinating. Nonetheless, not even one single experiment over the many performed over the decades has seen anything. Fortunately, due to the higher experimental control of condensed matter systems, it is now becoming increasingly popular to reproduce various aspects of fundamental physics in analog systems. Examples include the Hawking phenomenon in Bose--Einstein condensates~\cite{MunozdeNova2019}, the Weyl symmetry \cite{i2,weylgraphene} and the related Hawking/Unruh phenomenon on graphene~\cite{iorio2012,iorio2014,iorio2015}, gravitational and axial anomalies in Weyl semimetals~\cite{Gooth2017}, and more.

Henceforth, taking that road, we focus on the proposal of graphene as an analog of high-energy fundamental physics~\cite{i2,weylgraphene,iorio2012,iorio2014,iorio2015,ipfirst,CommentPRB2022,ip3,reach,GUPBTZ}, based on the fact that its low-energy excitations are massless Dirac pseudo-relativistic fermions (the matter fields $\psi$), propagating in a carbon two-dimensional honeycomb lattice~\cite{PacoReview2009}. Such behaviour is shared by a wide range of materials, called ``Dirac materials'' \cite{wehling}.

Here we shall rely heavily on \cite{threelayers} and, to a lesser extent, on \cite{ip2,ip5}, and shall take the chance to explain key details of the construction that are not explicitly exposed there.

\section{Variables, Supervariables, Hypervariables}
\label{section_var_sup_hyp}

In this problem we are in two-space and one-time dimensions. We have three energy ``layers'', each with its own appropriate phase-space variables, that we indicate as
\begin{equation}\label{layers}
  (x,p) \,, \quad (X_0, P_0) \,, \quad (X, P) \,.
\end{equation}
This notation is discussed at length in \cite{threelayers}. Let us say here that the \textit{variables} $(x,p)$ are the actual phase-space variables of the lowest energy Dirac quasiparticles of graphene that originate from the $O(\ell)$ dispersion relations, with $\ell$  the lattice spacing
\begin{equation}\label{lin dis rel}
   E_\pm = \pm v_F  |\vec{p}| \,,
\end{equation}
where $v_F = 3/2 \, |\eta_1| \, \ell/ \hbar$, with $\eta_1 \simeq - 2.8$eV the nearest neighbor hopping energy; see details in \cite{PacoReview2009}.

Such variables have a clear physical meaning, as coordinates and momenta of the low-energy conductivity electrons moving around (quantum mechanically) on the \textit{membrane}\footnote{This word is used on purpose, as the hexagonal lattice appears as a continuum at that energy/wavelength.}.

There we explain how to obtain the dispersion relations in the usual way (secular equation), based on the knowledge of the functions
\be
{\cal F}_m ({\vec{k}}) \equiv \sum_{i=1}^{n_m} e^{i {\vec{k}} \cdot {\vec{s}^{(m)}}_i} \,,
\ee
encoding the relevant information on the geometry of the lattice relative to the $m^{\rm th}$-near neighbors. Other details on the derivation of the ${\cal F}_m$s are in \cite{ip2}, where it is explained that the $n_m$ vectors $\vec{s}^{(m)}_i$ connect any given atom to its $m^{\rm th}$-near neighbors.

Using the fortuitous occurrence, ${\cal F}_2 = |{\cal F}_1|^2 - 3$, and various approximations explained in \cite{threelayers}, we arrive at the dispersion relations
\begin{equation}\label{disprel_1}
  E_\pm = \eta_1 (\pm |{\cal F}_1| - \tilde{A} |{\cal F}_1|^2) \;,
\end{equation}
where $\tilde{A} \simeq 0.15$.

It is then natural to define two further sets of variables, that we call ``\textit{supervariables}'', $(X_0, P_0)$, and ``\textit{hypervariables}'', $(X, P)$, as both sets refer to functions that appear in the dispersion relations that contain \textit{all orders}\footnote{This is literally true, as the function in (\ref{disprel_1}) contains all orders in $\ell$. Nonetheless, one needs to decide at which order one would like to proceed, say $O(\ell^m)$, and then all quantities need be taken at order $m$. Furthermore, the simple expression (\ref{disprel_1}) is already an approximation that holds when higher-order contributions are not considered, see \cite{threelayers,ip2,ip5}.} in $\ell$. In other words (writing things in a dimensionless fashion, i.e., not including a necessary factor of $1/\ell$)
\begin{equation}\label{3layersmomenta}
|{\cal F}_1| \sim p + O(\ell^2) \,, \quad |{\cal F}_1| \sim P_0 \,, \quad  |{\cal F}_1| - \tilde{A} |{\cal F}_1|^2 \sim P \,.
\end{equation}
These relations are key in our analysis and are dictated by the phenomenology of the system.

In \eqref{disprel_1}, if we define the supermomenta as
\begin{equation}\label{supermomenta}
\vec{P_{0}} = - \frac{\hbar}{\ell}(\mbox{Re}{\cal F}_1,\mbox{Im} {\cal F}_1) \;,
\end{equation}
then we can write
\begin{equation}\label{disprel_2}
 E_\pm = V_{F} (\pm |\vec{P_{0}}| - A\, |\vec{P_{0}}|^2) \,,
\end{equation}
where $V_{F} \equiv |\eta_{1}| \,\ell / \hbar$ and $A=\ell / \hbar \,\tilde{A}$.

Notice that in the following, we shall use a slightly different definition of $\vec{P_0}$, that allows one to expand it as $\vec{P_0} \sim \vec{p} + \cdots $, rather than $\vec{P_0} \sim 3/2 \vec{p} + \cdots $.
That is $\vec{{\cal P}_0} \equiv 2/3 \vec{P_0}$, but, to simplify notation with an abuse, we shall write $\vec{{\cal P}_0} \to \vec{P_0}$.

Let us now set $\hbar = 1$, unless otherwise stated.

Just like in the literature it is customary to use a standard Dirac Hamiltonian, in relation to the linear dispersion relations (\ref{lin dis rel}), we can keep going with the same logic by building-up the appropriate Hamiltonian in terms of the supermomenta $\vec{P_{0}}$, whose dispersion relations, descending
from the secular equation
\begin{equation}\label{Sec_Eq}
{\rm det} (H - E\mathbb{1}) = 0
\end{equation}
are indeed \eqref{disprel_2}. To do that one must realize that to obtain, through (\ref{Sec_Eq}), $|\vec{P_{0}}|$ in the dispersion relation, a corresponding term $\vec{\sigma} \cdot \vec{P_{0}}\equiv \slashed{P_{0}}$ needs be in the Hamiltonian:
$|\vec{P_{0}}| \to \slashed{P_{0}}$. Thus, the effective Hamiltonian we can use to describe the physics of the ``superlayer'' $(X_0,P_0)$ is
\begin{equation}
H (P_0) = V_{F}\,\sum_{k}\,\psi^{\dagger}_{k} \left(\slashed{P_{0}} - A \slashed{P_{0}} \slashed{P_{0}} \right) \psi_{k}
  = V_{F}\,\sum_{k}\,\psi^{\dagger}_{k} \left(\vec{\sigma} \cdot \vec{P_{0}} - A |\vec{P_{0}}|^2 \mathbb{1} \right) \psi_{k}  \,.\label{Hamiltonan_supermomenta}
\end{equation}
Notice that we use the convention\footnote{There are many conventions related to different choices for the pairs of inequivalent Dirac points, and different arrangements of the $a$ and $b$ operators to form the spinor $\psi$. See Appendix B of \cite{ip3} for more details.} $\psi_{k}^{\dagger}=(b_{k}^{\dagger},a_{k}^{\dagger})$, where $a_{k}$ and $b_{k}$ are the annihilation operators of each graphene sublattice, see \cite{i2}.

Indeed, by considering the $2 \times 2$ structure of $H$
\begin{equation}\label{H_blocks}
  H (P_0) \sim V_F \left(
              \begin{array}{cc}
                - A |\vec{P_{0}}|^2 &  P_0^1 - i P_0^2 \\
                P_0^1 + i P_0^2 & - A |\vec{P_{0}}|^2  \\
              \end{array}
            \right) \;,
\end{equation}
clearly the two solutions of (\ref{Sec_Eq})
\begin{equation}\label{H_blocks}
{\rm det} (H (P_0) - E \mathbb{1}) = 0 \sim {\rm det} \left(
              \begin{array}{cc}
                - A |\vec{P_{0}}|^2 - E &  P_0^1 - i P_0^2 \\
                P_0^1 + i P_0^2 & - A |\vec{P_{0}}|^2 - E \\
              \end{array}
            \right) = 0 \sim (A |\vec{P_{0}}|^2 + E)^2 = |\vec{P_{0}}|^2 \;,
\end{equation}
are those given in (\ref{disprel_1}).

We can then move one last step up, in energy and symmetry (see \cite{threelayers}), by defining the hypermomenta
\begin{equation}\label{hypermomenta}
\vec{P}=\vec{P_{0}}( 1 - A \, |\vec{P_{0}}|) \;,
\end{equation}
that is nothing more than a consequence of the above mentioned fortuitous occurrence, ${\cal F}_2 = |{\cal F}_1|^2 - 3$, hence a physical fact of this system; see (\ref{3layersmomenta}) and discussion there.

With this definition we arrive at the most symmetric situation of all, that is a full linearization of the Hamiltonian
\begin{equation}
H (P) = V_{F}\,\sum_{k}\,\psi^{\dagger}_{k} \, \slashed{P} \, \psi_{k} \,,\label{Hamiltonan_hypermomenta}
\end{equation}
with dispersion relations
\begin{equation}\label{lin dis rel_P}
   E_\pm = \pm V_F  |\vec{P}| \,,
\end{equation}
which is, in all respects but two, similar to (\ref{lin dis rel}). The two differences are crucial, though. First, the slope here is $V_F$ not $v_F  \simeq 1.5 V_F$. Second, the hypermomenta are generalized variables, containing, in principle, all orders in $\ell$ and $p$ (see \cite{threelayers} and later here).

To obtain $H(P_0)$ from $H(P)$ one should first use the expansion (\ref{hypermomenta}), and then use the prescription $|\vec{P_{0}}| \to \slashed{P_{0}}$
\begin{eqnarray}
H (P(P_0)) & = & V_{F}\,\sum_{k}\,\psi^{\dagger}_{k} \, \vec{\sigma} \cdot \vec{P} \, \psi_{k} \nonumber \\
           & = & V_{F}\,\sum_{k}\,\psi^{\dagger}_{k} \vec{\sigma} \cdot \left(\vec{P_{0}}(1 - A |\vec{P_{0}}|)\right) \, \psi_{k} \nonumber \\
           &\to& V_{F}\,\sum_{k}\,\psi^{\dagger}_{k} \left( \vec{\sigma} \cdot \vec{P_{0}} - A (\vec{\sigma} \cdot \vec{P_{0}}) \, (\vec{\sigma} \cdot \vec{P_{0}}) \right) \, \psi_{k} \nonumber \\
           & = & V_{F}\,\sum_{k}\,\psi^{\dagger}_{k} \left(\slashed{P_{0}} - A |\vec{P_{0}}|^2 \right) \, \psi_{k} = H (P_0) \,.\label{Hamiltonan_hyper_supermomenta}
\end{eqnarray}

Now, let us move towards the measurable quantities of the lowest layer $(x,p)$. We first need to expand ${\cal F}_1$ in powers of $\ell$ or $p$. Indeed, this means, through \eqref{supermomenta}, to expand $P_{0}$ in powers of $\ell$ or $p$, and through \eqref{hypermomenta}, to expand $P$ in powers of $\ell$ or $p$. The function in point is
\begin{equation}\label{F_1}
{\cal F}_1 = e^{i \ell k_{y}}\,\left(1+2\,e^{i \frac{3}{2}\ell\,k_{x}}\,\cos(\frac{\sqrt{3}}{2}\ell\,k_{x})\right) \;.
\end{equation}
This expansion, in general, needs be done around the two inequivalent Fermi points $K^{D}_{\pm} = (\pm\frac{4\pi}{3\sqrt{3}\ell},0)$, located where ${\cal F}_1 = 0$. In our case, though, we are not considering topological defects or other physical set-ups that include, in their description, both Dirac points. For us it is sufficient to do all calculations near one Dirac point, e.g., if $k_{i} = K^{D}_{+ i} + p_{i}$.

Thus, the Hamiltonian $H(P_0)$ in (\ref{Hamiltonan_supermomenta}) expressed in terms of $\vec{p}$ up to\footnote{One should recall that, in the Hamiltonian, there is always an overall factor $O(\ell)$, given by the Fermi velocity $v_F = \frac{3 \eta_1}{2 \hbar}  \times \ell$, so that $O(p^m)$ and $O(\ell^m)$, in the Hamiltonian, go actually together.} $O(\ell^{3})$, is
\begin{equation}\label{Hamiltonian_total_Op3}
\begin{split}
H(P_0(p)) = H(p)  = & \ v_F \sum_{\vec{p}} \psi^\dag_{\vec{p}} \big[ \sigma_1 \left( p_1 - \frac{\ell}{4} (p^2_1 - p^2_2) - \frac{\ell^2}{8} p_1 (p^2_1 + p^2_2) \right) \nonumber \\
   + & \ \sigma_2 \left( p_2 + \frac{\ell}{2} p_1 p_2 - \frac{\ell^2}{8} p_2 (p^2_1 + p^2_2) \right) \nonumber \\
   - & \ \frac{3}{2} A \left( (p^2_1 + p^2_2) - \frac{\ell}{2} p_1^3 + \frac{3 \ell}{2} p_1 p^2_2 \right) \big]  \psi_{\vec{p}} \;.
  \end{split}
\end{equation}

For the sake of complete control of the units, let us now reintroduce $\hbar$, and let us write the dispersion relations for this Hamiltonian as
\begin{eqnarray}\label{DispRelsmallp}
  E_\pm  = v_F \left[ \pm |\vec{p}| \mp  \frac{1}{4} \frac{\ell}{\hbar} |\vec{p}|^2 \cos(3\theta) \mp  \frac{1}{64} \frac{\ell^2}{\hbar^2} |\vec{p}|^3 \left(7 + \cos(6 \theta)\right)
          -  \frac{3}{2} \, A \, |\vec{p}|^2 + \frac{3}{4} \, A \, \frac{\ell}{\hbar} |\vec{p}|^3 \cos(3\theta) \right] \;,
\end{eqnarray}
where
\begin{equation}\label{A}
  A = 0.15 \, \frac{\ell}{\hbar} > 0 \;,
\end{equation}
$\tan \theta = p_2/ p_1$ and
\begin{equation}\label{FermiVelocity}
 v_F \equiv \frac{3}{2} \, |\eta_1| \, \frac{\ell}{\hbar} \, \simeq c/300 \simeq 10^6 m/s \;.
\end{equation}

Let us give here another writing of the dispersion relations
\begin{eqnarray}
  E_\pm  & = & \frac{3}{2} |\eta_1| \left[ \pm \left(\frac{\ell}{\hbar} |\vec{p}| \right) \mp  \frac{1}{4} \left(\frac{\ell}{\hbar} |\vec{p}| \right)^2 \cos(3\theta) \mp  \frac{1}{64} \left(\frac{\ell}{\hbar} |\vec{p}| \right)^3 \left(7 + \cos(6 \theta)\right) \right. \nonumber \\
         & - & \left. \frac{3}{2} \, 0.15 \, \left(\frac{\ell}{\hbar} |\vec{p}| \right)^2 + \frac{3}{4} \, 0.15 \, \left(\frac{\ell}{\hbar} |\vec{p}| \right)^3 \cos(3\theta) \right] \label{DispRelsmallpExpPar} \;,
\end{eqnarray}
which makes clear that the expansion parameter is $\frac{\ell}{\hbar} |\vec{p}|$.

\section{Choice of the generalized coordinates and related GUPs}
\label{section_coordinates}

From the previous discussion, it is clear that the interpretation of $P^i_0$ and $P^i$ as generalized momenta is suggested by the dispersion relations, hence the functional form of $P_0^i(p)$
\begin{align}\label{SupermomentumCartesian}
P^1_0 & =  p^1 + \frac{\ell}{4} \left((p^2)^2 - (p^1)^2\right) -  \frac{\ell^2}{8} p^1 \left((p^1)^2 + (p^2)^2\right) \,, \nonumber \\
P^2_0 & =  p^2 + \frac{\ell}{2} p^1 p^2 -  \frac{\ell^2}{8} p^2 \left((p^1)^2 + (p^2)^2\right)  \,.
\end{align}
or, equivalently
\begin{align}\label{P_0 trigonal}
P^1_0 & =  p^1 - \frac{\ell}{4} |\vec{p}|^2 \cos 2 \theta -  \frac{\ell^2}{8} |\vec{p}| \cos \theta \;, \nonumber \\
P^2_0 & =  p^2 + \frac{\ell}{4} |\vec{p}|^2 \sin 2 \theta -  \frac{\ell^2}{8} |\vec{p}| \sin \theta \;,
\end{align}
Hence of $P^i(p)$, through (\ref{hypermomenta}), are dictated by the physics of the system. The form of the dispersion relations, in turn, tells us how the effective Hamiltonian must look like, with little or no freedom.

What is left to the interpretation is the meaning of the generalized coordinates, $X^i_0$ and $X^i$. Only when this is done the picture is complete, and we have all that is necessary, $(H, X, P)$, to describe the dynamics.

Although there is room for different interpretations for the generalized coordinates, there are two choices that are very natural, hence stand out as privileged:

\begin{itemize}
  \item The first choice is to keep $x^i$ as the generalized coordinates, $x^i = X_0^i = X^i$. This choice, as obvious from $[x, P(p)] \sim i \partial_p P$, leads to some form of GUP.
  \item The second choice is to demand the generalized coordinates to be canonical, $[X_0, P_0] = i = [X, P]$. This choice necessarily leads to both $X_0^i(x,p)$ and $X^i(x,p)$.
\end{itemize}

To summarize (see \cite{threelayers} for details): One can move from the standard (canonical) variables that describe the lowest energy physics of the Dirac quasi-particles of graphene, $(x,p)$, to the next ``layer'', by including higher order contributions in $p$ to $P_0$, that we call supervariables, $(X_0,P_0)$. Then, considering that contributions of the same order could come by including next-to-near neighbors in the computations ($A \neq 0$), one can move one more ``layer'' up, to the hypervariables $(X,P)$.

If, at every step, we only consider the two privileged choices for the generalized coordinates, the GUP and the canonical, the overall options we have are given in the following diagram \cite{threelayers}
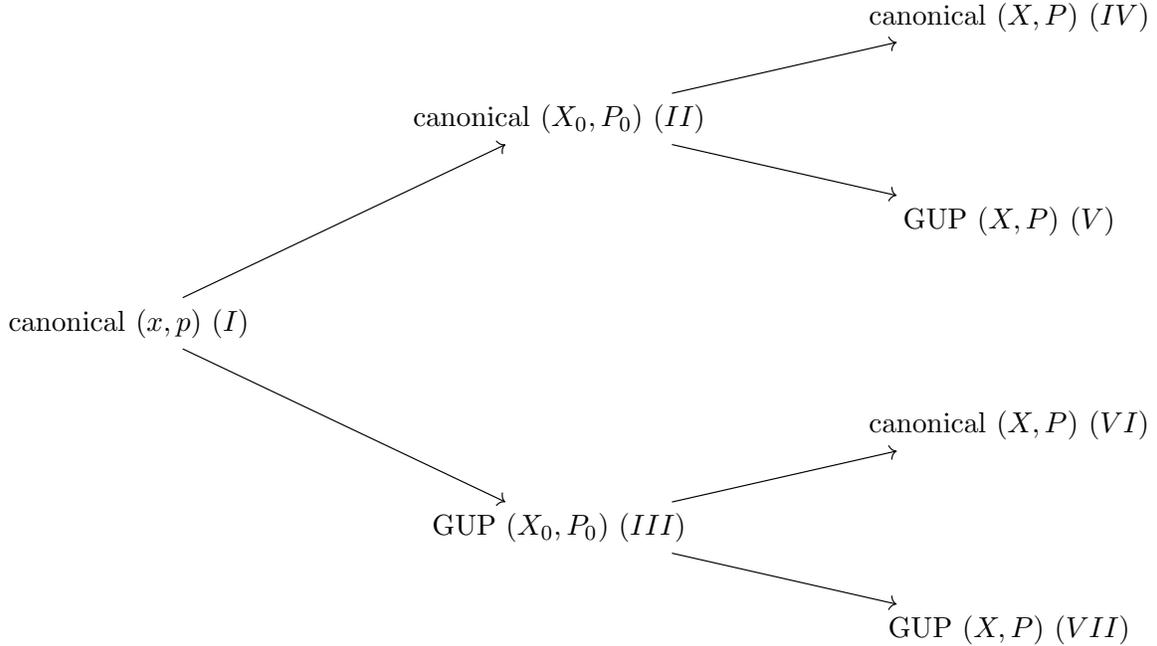
\begin{equation*}
\begin{tikzcd}[column sep=huge]
                                                   &                                                       & \mbox{canonical $(X,P)$ $(IV)$}  \\
                                                   &   \mbox{canonical $(X_0,P_0)$ $(II)$} \ar[ur] \ar[dr] &                                  \\
                                                   &                                                       & \mbox{GUP $(X,P)$ $(V)$}         \\
\mbox{canonical $(x,p)$ $(I)$}  \ar[uur] \ar[ddr]  &                                                       &                                  \\
                                                   &                                                       & \mbox{canonical $(X,P)$ $(VI)$}  \\
                                                   &   \mbox{GUP $(X_0,P_0)$ $(III)$} \ar[ur] \ar[dr]      &                                  \\
                                                   &                                                       & \mbox{GUP $(X,P)$ $(VII)$}
\end{tikzcd}
\end{equation*}

It is important to notice that:

\begin{itemize}
  \item To include terms from $O(p^2)$ on, we have to keep $A \neq 0$, as seen in (\ref{A}). Therefore, the ``layer of reference'' is always the ``hyperworld'', $(X,P)$.
  \item The interesting set-ups are those leading to GUPs in the final layer, the one of reference. In the above scheme, these are (V) and (VII).
  \item No matter the choice, all the quantities eventually need be expressed in terms of the measurable standard variables, $(x,p)$.
\end{itemize}

With this in mind, the GUP-corrected physics here is given by the Hamiltonian (\ref{Hamiltonian_total_Op3}), and the related dispersion relations (\ref{DispRelsmallp}). We simply need to read-off all results in terms of the appropriate variables, that for the coordinates are:

\begin{enumerate}
  \item the case $X^i = x^i$ (that is the ladder $(I) \to (III) \to (VII)$), leading to the GUP (VII)
\begin{equation}
[X^{i},P^{j}] =  \ii F^{i\,k} (\vec{P})\, \left[ \delta^{kj}
- A   |\vec{P}|   \, \left( \delta^{kj} + \frac{P^{k}\,P^{j}}{|\vec{P}|^2}\right)
- A^2 |\vec{P}|^2 \, \left( \delta^{kj} + \frac{P^{k}\,P^{j}}{|\vec{P}|^2}\right) \right]
\end{equation}
where
\begin{equation*}
  F^{i\,j}(\vec{P}) = \delta^{i j}
  + \frac{1}{2} \, \ell \, \left(
                  \begin{array}{cc}
                  - P^1 & P^2 \\
                    P^2 & P^1 \\
                  \end{array}
                \right)
  + \frac{1}{2} \, \ell \, A \, |\vec{P}| \, \left(
                  \begin{array}{cc}
                  - P^1 & P^2 \\
                    P^2 & P^1 \\
                  \end{array}
                \right)
  - \frac{1}{2} \, \ell^2 \, \left(
                  \begin{array}{cc}
                    (P^1)^2  & P^1 P^2 \\
                    P^1 P^2 &  (P^2)^2   \\
                  \end{array}
                \right) \;,
\end{equation*}
Notice that $F^{ij}  = F^{ji}$, and $F^{i}_i (\vec{P}) = 2 - \frac{\ell^2}{2} |\vec{P}|^2$.

  \item and the case $X^i = X^i_0 (x,p)$ with
  \begin{align} \label{supercoordinates_canonical}
X_{0}^{1} (x,p) &= \left[1 + \frac{\ell}{2} p^{1} + \frac{\ell^{2}}{8}\left( 5 (p^{1})^{2} + 3 (p^{2})^{2} \right)\right] \, x^{1} + \left[-\frac{\ell}{2} p^{2} + \frac{\ell^{2}}{4} p^{1} p^{2}\right] \, x^{2}  \;,  \\
X_{0}^{2} (x,p)&= \left[-\frac{\ell}{2} p^{2} + \frac{\ell^{2}}{4} p^{1} p^{2} \right] \, x^{1} + \left[1 - \frac{\ell}{2} p^{1} + \frac{\ell^{2}}{8}\left( 3 (p^{1})^{2} + 5 (p^{2})^{2} \right) \right] \, x^{2}  \;. \nonumber
\end{align}
That is the ladder $(I) \to (II) \to (V)$), leading to the GUP (V), an expression coinciding, at first order in $A$, with the ADV-like GUP (see (\ref{ADVGUPtrue}))
\begin{equation} \label{ADVGUP}
[X^{i},P^{j}] =  \ii \,\left[ \delta^{ij}
- A   |\vec{P}|   \, \left( \delta^{ij} + \frac{P^{i}\,P^{j}}{|\vec{P}|^2}\right)
- A^2 |\vec{P}|^2 \, \left( \delta^{ij} + \frac{P^{i}\,P^{j}}{|\vec{P}|^2}\right) \right] \;.
\end{equation}
\end{enumerate}
The other commutators are trivial.

One does not need any special or exotic setup to have the GUP all-the-way ladder at work. The only thing that is necessary is to include higher order terms in the dispersion relations, and then move to the next-to-near neighbors. The coordinates are always just $x^i$. Since the literature is full of proposals on how GUPs affect a variety of phenomena, see, e.g., \cite{Maggiore:1993rv,Kempf:1993bq,Scardigli:1999jh,Buoninfante:2019fwr,Petruzziello:2020wkd,Bosso:2021koi,GUPBTZ} and since we have obtained here many such GUPs, some very similar to the ones descending from quantum gravity, then we are now in the position to prove some of those theoretical conjectures in the lab.

Notice that, in going from $(I)$ to $(III)$, we also have a GUP
\begin{equation}\label{Super Noncommutative Snyder}
  [X_0^i , P_0^j] = \ii F^{i j} (\vec{P_0}) , \ \ \ [X_0^i , X_0^j] =  0 = [P_0^i , P_0^j]  \;,
\end{equation}
with $F^{i j} (\vec{P_0})$ given in
\begin{equation*}
  F^{i\,j}(\vec{P_0}) = \delta^{i j}
  + \frac{1}{2} \, \ell \, \left(
                  \begin{array}{cc}
                  - P_0^1 & P_0^2 \\
                    P_0^2 & P_0^1 \\
                  \end{array}
                \right)
  - \frac{1}{2} \, \ell^2 \, \left(
                  \begin{array}{cc}
                    (P_0^1)^2  & P_0^1 P_0^2 \\
                    P_0^1 P_0^2 &  (P_0^2)^2   \\
                  \end{array}
                \right) \;.
\end{equation*}
Since, for the reasons explained earlier, we shall always like to include the $A\neq0$ effects, we only consider this as an intermediate mathematical step.

\section{Conclusions}

We have discussed here how analogs of three kinds of GUPs can be obtained when higher order terms are considered in the dispersion relations of the conductivity electrons of a generic Dirac material.

An analog system cannot behave as the target system in all respects but only in certain specific conditions, for certain types of particles and for particular regimes of the interaction. In fact, one of the aspects we have to give up in an analog is the \emph{universality}, which should be expected in the presence of a \emph{fundamental} GUP.

For instance, in the ``three graphene worlds'' we have explored here, we only have at our disposal a limited set of all the possible particles and fields, compared to those that nature gives us. Essentially, in a Dirac material we have analogs of
\begin{itemize}
  \item massless Dirac spinors, $\psi$, that are the $\pi$-electrons quasiparticles;
  \item the U(1) gauge fields, $A_\mu$, stemming from straining the material, that could be used to mimic electromagnetic interactions;
  \item the nonabelian gauge fields, $A^a_\mu$, due to the two copies of $\psi$, one per Dirac point, often described as SU(2) gauge fields, that could be used to mimic internal symmetries;
  \item the spin connection, $\omega^a_\mu$, that always takes care of the intrinsic curvature associated to disclination defects in the lattice, and may or may not include contorsion, $\kappa^a_\mu$, that takes care of the intrinsic curvature associated to dislocation defects;
  \item classical (and quantum) metric fields, $g_{\mu \nu}$, that emerge as a collective description of the elastic and unelastic properties of the membrane.
\end{itemize}
Some more steps could be moved when we adapt (reduce) external fields, e.g., the true electromagnetic field, to the lower dimensional, and $v_F$-relativistic (rather than the $c$-relativistic) and specific dynamics of the system. That is all.

Clearly, then, here we cannot have all types of particles and fields propagating in a GUP-compatible space, described by a suitable $g_{\mu \nu}$, but we can do that for a limited set, for instance, $\psi$ propagating on a flat space with granular structure.

Even for this limited situation, since the recipe here consists in having control on the appropriate phase-space variables, we need to know how to recognize the ``shadows of the high energy physics'', so to speak, on what mimics ``the low energy GUP-corrected physics''.

With the results of \cite{threelayers}, we are now able to face these issues because we now know the form of $X^i$ and $X_0^i$, in terms of $x^i$ and $p^i$. With those results in our hands, we may safely solve the problem of lack of universality by, e.g., limiting the set-up to only Dirac quasi-particles $\psi$ and setting up a potential well, as many others do routinely in the same analog/effective language.

Such potential $V$ must be a function of the correct variables, $X^i$, i.e. $V(X)$. But now we know $X(x,p)$, and this simply means $X_i = x_i$ in the most interesting case. Hence, the very same potential well $V(x)$ one has for the GUP-non-corrected physics (the standard one) stays for GUP-corrected physics. In an experiment like that, we would prove that the GUP-corrected dynamics can indeed be practically realized in such a simple system, and one could check many theoretical statements in a lab.

To close, a note on the generality of our results. All condensed matter systems have corrections to linear or quadratic dispersion laws. Some kind of GUP may then appear in many systems after the suitable variables change. Notice that our approach may or may not lead to sensible analogs GUPs.  In other words, it is not as simple as ``any suitable change of variables leads to some sensible GUP''. For instance, in our case, we have seen that only by stopping at ${\cal F}_2$ and making other approximations, it is possible to find GUPs related to the ADV GUPs. If one keeps going in the order of the expansion, no natural ADV or ADV-like structures are recovered. Other systems may have different natural GUPs, none at all, or unknown GUPs (like some of those that we discussed here).

To close, let us stress again that, of course, we did not prove here that graphene, or other materials of that kind, are systems where \textit{fundamental} GUPs are at work. Nor we are suggesting that, following our procedure, fundamental GUPs can be spotted in other condensed matter systems. We are saying here that those easy-to-produce tabletop real systems, by a simple redefinition of the dynamical variables, can be turned into very efficient analogs of the fundamental scenarios that are supposed to be responsible for the GUPs that the high energy physics community is eagerly looking for. This seems promising, especially in the light of how simple our recipe is and of how difficult it is to run high energy experiments.

\section*{Acknowledgements}
A.~I. and P.~P. sincerely thank Thomas Elze and the other organizers of DICE2022, for a wonderful and very productive conference. They gladly acknowledge the financial support of Charles University Research Center (UNCE/SCI/013). P.~P. is also supported by Fondo Nacional de Desarrollo Cient\'{i}fico y Tecnol\'{o}gico--Chile (Fondecyt Grant No.~3200725). S.~M. acknowledges support from COST action CA18108.

\bibliography{dice2022_three_layers_biblio}{}
\bibliographystyle{iopart-num}

\end{document}